\documentclass{vgtc}                          

\graphicspath{{figures/}{pictures/}{images/}{./}} 

\usepackage{times}                     

\usepackage{tabu}                      
\usepackage{booktabs}                  
\usepackage{lipsum}                    
\usepackage{mwe}                       

\usepackage{mathptmx}                  
\usepackage{color}

\onlineid{0}

\vgtccategory{Research}

\vgtcinsertpkg

\usepackage{orcidlink}

\title{Exploring the Impact of Passthrough on VR Exergaming in Public Environments: A Field Study}

\author{Zixuan Guo \orcidlink{0000-0002-0451-8988}\\ %
       \parbox{1.4in}{\scriptsize \centering Xi'an Jiaotong-Liverpool University \\ The University of Liverpool} %
\and Hanxiao Deng \orcidlink{0009-0005-4587-550X}\\ %
    \parbox{1.4in}{\scriptsize \centering Xi'an Jiaotong-Liverpool University} %
\and Hongyu Wang \orcidlink{0000-0002-2288-5116}\\ %
    \parbox{1.4in}{\scriptsize \centering Xi'an Jiaotong-Liverpool University} %
\and Angel Tan \orcidlink{0000-0002-6947-3063}\\ %
    \parbox{1.4in}{\scriptsize \centering Aston University} %
\and Wenge Xu \orcidlink{0000-0001-7227-7437}\\ %
    \parbox{1.4in}{\scriptsize \centering Birmingham City University} %
\and Hai-Ning Liang~\orcidlink{0000-0003-3600-8955}\thanks{Corresponding author (e-mail: hainingliang@hkust-gz.edu.cn)}\\ %
    \parbox{1.4in}{\scriptsize \centering The Hong Kong University of Science and Technology (Guangzhou)}}

\author{
Zixuan Guo\textsuperscript{1}\textsuperscript{2}~\orcidlink{0000-0002-0451-8988}
\and 
Hanxiao Deng\textsuperscript{1}~\orcidlink{0009-0005-4587-550X}
\and 
Hongyu Wang\textsuperscript{1}~\orcidlink{0000-0002-2288-5116}
\and 
Angel J. Y. Tan\textsuperscript{3}~\orcidlink{0000-0002-6947-3063}
\and 
Wenge Xu\textsuperscript{4}~\orcidlink{0000-0001-7227-7437}
\and 
Hai-Ning Liang\textsuperscript{5}~\orcidlink{0000-0003-3600-8955}\thanks{Corresponding author (e-mail: hainingliang@hkust-gz.edu.cn)}
}

\affiliation{\scriptsize \textsuperscript{1} School of Advanced Technology, Xi'an Jiaotong-Liverpool University, Suzhou, China\\
\textsuperscript{2} Department of Computer Science, The University of Liverpool, Liverpool, UK\\
\textsuperscript{3} Department of Psychology, Aston University, Birmingham, UK\\
\textsuperscript{4} College of Computing, Birmingham City University, Birmingham, UK\\
\textsuperscript{5} Computational Media and Arts Thrust, The Hong Kong University of Science and Technology (Guangzhou), Guangzhou, China}

\teaser{
  \centering
  \includegraphics[width=\linewidth]{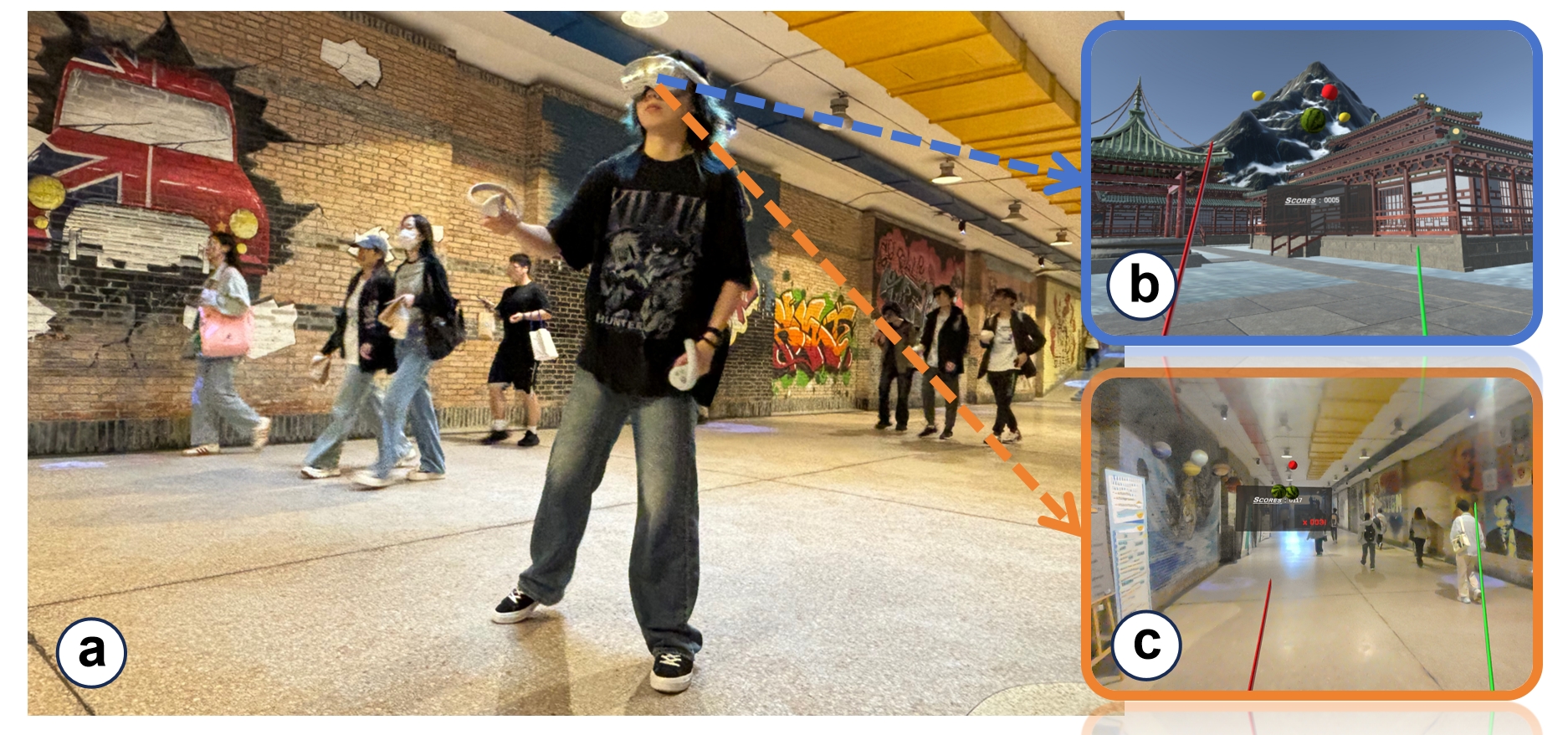}
  \caption{(a) A user is playing VR exergames in a public environment while other passersby are walking by. (b) Virtual view, displaying the complete virtual environment. (c) Passthrough view, enabling environmental awareness during gameplay.}
  \label{fig:teaser}
}

\abstract{
Sedentary behavior is becoming increasingly prevalent in daily work and study environments. VR exergaming has emerged as a promising solution in these places of work and study. However, private spaces in these environments are not easy, and
engaging in VR exergaming in public settings presents its own set of challenges (e.g., safety, social acceptance, isolation, and privacy protection). The recent development of Passthrough functionality in VR headsets allows users to maintain awareness of their surroundings, enhancing safety and convenience. Despite its potential benefits, little is known about how Passthrough could affect user performance and experience and solve the challenges of playing VR exergames in real-world public environments. To our knowledge, this work is the first to conduct a field study in an underground passageway on a university campus to explore the use of Passthrough in a real-world public environment, with a disturbance-free closed room as a baseline. Results indicate that enabling Passthrough in a public environment improves performance without compromising presence. Moreover, Passthrough can increase social acceptance, especially among individuals with higher levels of self-consciousness. These findings highlight Passthrough's potential to encourage VR exergaming adoption in public environments, with promising implications for overall health and well-being.

} 

\keywords{Virtual reality, exergaming, passthrough functionality, public environments, gameplay mechanics.}

\begin{document}


\firstsection{Introduction}

\maketitle

In today's society, many individuals sit for extended periods due to work or study requirements, leading to insufficient physical activity. Studies \cite{castro2020sedentary, clemes2014office, edelmann2022physical} show high levels of sedentary behavior among office workers and students. Office workers spend up to 71\% of their workday seated, and university students average 7.29 hours of sedentary time daily \cite{castro2020sedentary, clemes2014office}. Research \cite{akksilp2023physical} suggests that incorporating multiple brief (i.e., at least 4 minutes) physical activity sessions during long periods of sitting can improve physical health. However, due to barriers like lack of motivation, fatigue, and time constraints, 27.5\% of adults struggle to meet the World Health Organization's recommendation of at least 150 minutes of moderate or 75 minutes of vigorous physical activity per week \cite{bull2020world}.

Virtual Reality (VR) exergames offer a fun and immersive way to combine physical activity with virtual environments, breaking the monotony of traditional workouts. Studies \cite{yoo2020embedding, touloudi2022applicability, stranick2021virtual, xu2021effects, sousa2022active, liu2021effects} have highlighted their benefits for sedentary workers and students. For example, Yoo et al. \cite{yoo2020embedding} provided a closed room for sedentary workers to exercise during breaks, resulting in significant physical activity and mood benefits. However, setting up such rooms can be costly and impractical due to space limitations. Playing VR exergames in public shared environments is a more cost-effective solution.

Public environments, despite being low-cost and accessible, are dynamic and uncontrolled, posing challenges for VR users. Unpredictable foot traffic and the presence of passersby can create social obstacles and safety concerns \cite{mai2018public, eghbali2019social, gugenheimer2019challenges}. However, the introduction of full-color Passthrough in recent devices (e.g., Meta Quest 3, Pico 4, Apple Vision Pro) allows VR users to maintain awareness of their physical surroundings. Studies suggest that Passthrough enhances safety and situational awareness, reducing collision risks \cite{o2023re, o2022exploring}. Additionally, it allows interaction with bystanders and the environment without removing the headset, improving convenience and interaction flow \cite{o2023re, wang2022realitylens, guo2024enhancement}. 

Existing literature \cite{guo2024breaking, o2023re, von2019you, mcgill2015dose, wang2022realitylens} only investigates the Passthrough feature in controlled laboratory settings, where bystanders are portrayed by trained experimenters performing specific tasks such as observation, conversation, and interruption. There is a significant research gap in the use of the Passthrough feature in (1) a natural set-up, where bystanders continue to their daily routine/task rather than performing a given task, and (2) a public environment, which is dynamic and more complex than a controlled laboratory environment, raising doubts and questions about the applicability of these research findings to real-world public environments. Given these research gaps, conducting field studies in real-world public environments to explore how Passthrough influences users' performance and experience in VR exergaming is timely and necessary. 

To the best of our knowledge, our work is the first to explore Passthrough in real-world public environments, investigating three conditions: Closed Room (Baseline; in a small room without distractions \cite{yoo2020embedding}), Public Environment (PE), and Public Environment with Passthrough (PE-P). The public environment used in our experiment was an underground passageway on a campus, with stable yet moderate foot traffic, making it suitable for VR exergaming and representative of public settings like malls and parks. To enable participants to maintain environmental awareness while engaging in VR exergaming, we used Passthrough Augmented Reality (PAR) \cite{guo2024breaking, o2023re}, preserving key gaming elements and overlaying the rest with the Passthrough view.

Our findings indicate that while participants performed worse in a public environment compared to a closed room, the use of Passthrough eliminated this social inhibition. Contrary to past research suggesting Passthrough might disrupt VR presence, our results showed no significant impact in public settings. Furthermore, Passthrough improved social acceptance and had a favorable impact on participants with higher self-consciousness. Given that VR exergaming has been shown to improve the health of sedentary people \cite{yoo2020embedding, touloudi2022applicability, stranick2021virtual, xu2021effects, sousa2022active, liu2021effects}, our findings suggest that Passthrough can facilitate VR integration into public environments where people work or study, addressing challenges and enhancing daily life.


\section{Related Work}
\subsection{Exercising Through VR Exergames in Daily Work or Study}
People spend a significant amount of time each day in work and study settings, and sedentary behavior has become increasingly common in daily life, leading to various detrimental health effects \cite{park2020sedentary}. Incorporating physical activity into work or study, even for multiple 4-minute sessions, has been proven to mitigate this issue \cite{osei2005effects, akksilp2023physical}. VR exergames have emerged as a promising avenue as they can make physical activities enjoyable and engaging, thereby increasing motivation for physical activity. Research indicates that compared to traditional forms of physical activities, VR exergames can promote better self-efficacy, active engagement, and enjoyment, and alleviate symptoms of depression \cite{mocco2024enhancing}.

Several studies \cite{yoo2020embedding, touloudi2022applicability, stranick2021virtual, xu2021effects, liu2021effects, sousa2022active} have explored the impact of using VR exergames for physical activities in the workplace and on campuses. Yoo et al. \cite{yoo2020embedding} explored this among 11 sedentary workers and found that engaging in VR games during work provided them with physical and mood benefits. Touloudi et al. \cite{touloudi2022applicability} reported favorable acceptance and enjoyment of VR exergames among 40 middle-aged female workers, who showed a positive attitude towards long-term use. Similarly, studies on students have shown positive effects of VR exergames on campus. Xu et al. \cite{xu2021effects} found high acceptance and potential depression reduction among 31 university students over six weeks. Liu \cite{liu2021effects} conducted a four-week study with 36 students, finding improved exercise motivation and mood states.

These studies demonstrate that VR exergames can provide an enjoyable way to motivate and interest people in physical activities within their daily work and study routines, thereby contributing to improving their physical health. However, it is worth noting that these work or study environments are often public spaces,  while Yoo et al. \cite{yoo2020embedding} used a small closed room within a workplace, which can be costly and has space constraints. In most cases, people need to use VR in public environments, where many others are present, posing a series of challenges.

\subsection{Challenges of VR Usage in Public Environments}
Public environments are characterized by their open spaces and continuous flow of people, making them dynamic and uncontrollable settings \cite{hamilton2008shared}. As VR headsets become increasingly portable, they unlock new avenues for integrating VR technology into public settings. However, this expansion also brings forth its own set of challenges. Safety is a primary concern in these environments, where users may have reduced awareness of their surroundings, increasing the risk of collisions or falls \cite{hartmann2019realitycheck}. Additionally, VR users may experience unintended collisions with bystanders, posing risks to both parties \cite{dao2021bad}.


These public environments also pose social challenges, including issues related to social acceptance, isolation, and privacy protection. The noticeable appearance of VR equipment may draw unwanted attention or scrutiny from others, leading to feelings of self-consciousness or embarrassment among users, thereby reducing their acceptance of using VR in public environments \cite{schwind2018virtual, eghbali2019social, vergari2021influence, medeiros2023surveying}. Furthermore, VR headsets create a barrier between users and their surroundings, potentially hindering their ability to interact effectively and comfortably with bystanders, thereby placing them in a socially isolated position \cite{o2023re, o2021safety, dao2021bad, bajorunaite2023reality}. Additionally, privacy considerations are critical when deploying VR technology in public environments. The use of VR headsets in such spaces raises concerns about users being recorded without their consent by bystanders or malicious actors \cite{o2023privacy, o2023re}.

These challenges are even more pronounced for those with high self-consciousness. Self-consciousness refers to individuals' awareness of their thoughts, feelings, and behaviors in relation to others \cite{fenigstein1975public}. Those with higher levels of self-consciousness tend to have heightened concerns about social evaluation, leading to increased anxiety and decreased performance in social situations \cite{wang2004self}. Recent studies \cite{guo2024breaking, woods2022impact} underscore the significance of considering personality differences in the impact of using VR in public environments, particularly highlighting the role of self-consciousness.

In short, public environments pose significant challenges for the use of VR due to their dynamic and uncontrollable nature. Many studies \cite{o2021safety, eghbali2019social, mai2018public, bajorunaite2021virtual, williamson2019planevr} have highlighted people's concerns about using VR in public environments, which are heightened for individuals with high self-consciousness. Thus, addressing these issues is essential to foster a more responsible and inclusive use of VR in public environments. Many studies \cite{medeiros2021promoting, wu2023investigating, o2023re, von2019you} aim to enhance users' awareness of reality, and the most notable method among them is the Passthrough function, as it provides a real view of the physical world.

\subsection{Passthrough Functionality and Its Benefits}
The Passthrough functionality aimed at enhancing the user experience by breaking the isolation typically associated with VR headsets \cite{guo2024breaking, o2022exploring}. This feature enables users to view the real world while wearing the headset, achieved through either full Passthrough mode or switching to a Passthrough AR version (PAR) while using applications \cite{o2023re}. PAR involves preserving essential virtual elements while overlaying the remaining content with the Passthrough view \cite{guo2024breaking}. By leveraging built-in cameras to capture surroundings and display them in real-time, Passthrough seamlessly integrates the virtual and real worlds, resulting in a cohesive and immersive user experience \cite{luo2023achromatic}.

Passthrough offers the dual advantages of enhancing safety and situational awareness while facilitating a seamless transition between the virtual and real worlds. By allowing users to maintain awareness of their real-world environment while immersed in VR, Passthrough reduces the likelihood of collisions or hazards \cite{o2023re}. This heightened awareness is particularly valuable in shared or public settings, helping users avoid unintended interactions with bystanders \cite{guo2024breaking, o2023re, o2022exploring}. Additionally, Passthrough allows users to handle real-world tasks and interact with their environment and bystanders without removing the VR headset, ensuring an uninterrupted flow of the VR experience \cite{o2023re, wang2022realitylens}. This functionality also improves overall user experience and comfort during VR sessions by reducing disorientation and providing a smoother transition back to reality \cite{pointecker2022bridging}.

These advantages of the Passthrough feature have significant potential to help VR overcome challenges in public environments; however, there is currently a lack of field studies conducted in real public environments to explore its impact on user performance and experience. Previous research \cite{guo2024breaking, o2023re, von2019you, mcgill2015dose, wang2022realitylens} focusing on Passthrough often used controlled laboratory environments with trained experimenters acting as bystanders. For example, Guo et al. \cite{guo2024breaking} created controlled office and corridor environments with trained experimenters acting as bystanders to observe participants. Similarly, Willich et al. \cite{von2019you} had experimenters simulate bystanders randomly appearing in different positions around the participants. O'Hagan et al. \cite{o2023re} assessed the usability of the Passthrough feature by having participants imagine various scenarios in public environments, such as facing a crowd or someone with a pet, through the ``Wizard of Oz" method. While these studies provide valuable insights into understanding Passthrough's effects on user performance and experience, public environment situations are more complex and uncontrolled, and its use in real-world public environments has not yet been studied. Therefore, we conducted a field study in this paper to address these research gaps.

\section{Experiment}
\subsection{Experiment Design}
Given that our aim was to explore the impact of Passthrough on VR exergaming in public environments, it was essential for the public setting in our experiment to be representative. Guo et al. \cite{guo2024breaking} compared an office and a corridor, finding that participants prefer VR exergaming in corridor-like spaces because of the ample room available for their movements and reduced prolonged observation from moving passersby. For public environments with foot traffic, we considered the following factors: (1) they should be a public space suitable for physical activity; (2) they should have regular foot traffic to ensure all participants experience similar conditions; and (3) they should not be overly busy, allowing enough space for both users and passersby.

\begin{figure}[htbp]
    \centering
    \includegraphics[width=\columnwidth]{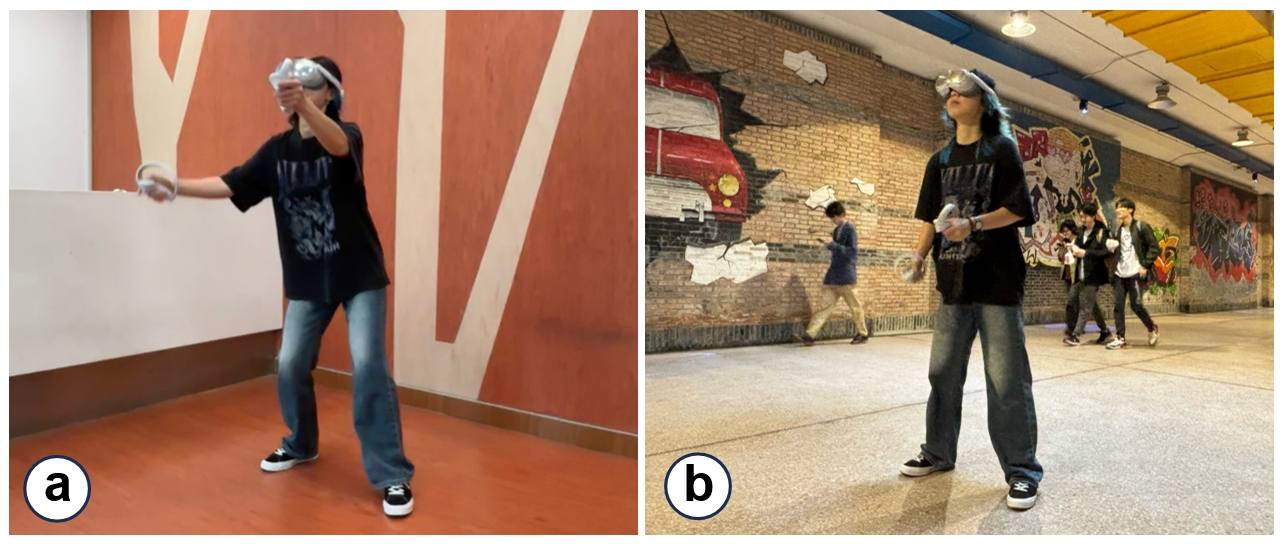}
    \caption{ A participant is engaging in VR exergaming in (a) a distraction-free closed room and (b) an underground passageway with continuous traffic.}
    \label{space}
\end{figure}

Thus, we selected an underground passageway within a university campus as our public environment (Figure \ref{space}). This location maintains a steady flow of foot traffic on weekdays without becoming overcrowded, making it representative of typical public settings such as shopping malls, high streets, and parks. Moreover, we employed a small closed room (Figure \ref{space}) to serve as the baseline condition for our study, providing an environment devoid of any disruptions \cite{yoo2020embedding}. Consequently, the experiment, which followed a within-subjects design, comprised three conditions that were counterbalanced. Here are the specifics:

\begin{itemize}
\item \textbf{Closed Room (Baseline):} Participants engaged in VR exergaming sessions in a small closed room with no disruptions.
\item \textbf{Public Environment (PE):} Participants engaged in VR exergaming sessions in the underground passageway.
\item \textbf{Public Environment with Passthrough (PE-P):} Participants engaged in VR exergaming sessions in the underground passageway with Passthrough functionality (PAR) enabled, allowing them to see their physical surroundings.
\end{itemize}

\subsection{Apparatus and Setup}
The experiment utilized a Pico 4 as the VR device and a Polar OH1 for tracking participants' heart rate and calorie expenditure. The experiment was conducted on weekdays, avoiding peak student class times to ensure participants experienced a similar public environment set up in the underground passageway. During most times, the passageway maintained a steady flow of foot traffic, with approximately 150 people passing through every 10 minutes. For the Baseline condition, a small closed room on campus, accessible within a 5-minute walk from the underground passageway, was utilized. Both environments were adequately lit to ensure participants could clearly see the game objectives regardless of Passthrough functionality. The indoor environment was set to 21°C throughout the experiment via central air conditioning, matching the average temperature of the underground passageway during the experiment. In all conditions, participants were under the observation and supervision of an experimenter to ensure their safety. This study obtained ethical approval from the University Ethics Committee and permission for site use from the Estate Management Department.

\subsection{VR Exergame}
We developed a game similar to VR Fruit Ninja \footnote{https://store.steampowered.com/app/486780/Fruit\_Ninja\_VR/} using the Unity3D engine, version 2021.3.26f1. VR Fruit Ninja has been utilized in many studies concerning exergaming \cite{guo2024breaking, yoo2020embedding, guo2023s} due to its ability to not only encourage player movement but also its straightforward gameplay that appeals to a diverse range of players. Following the game used by \cite{guo2024breaking}, we also introduced obstacle elements to the gameplay to encourage players to squat, thereby amplifying the overall physical activity level. The parameters mentioned below underwent refinement through extensive playtesting involving 3 testers.

\begin{figure}[htbp]
    \centering
    \includegraphics[width=\columnwidth]{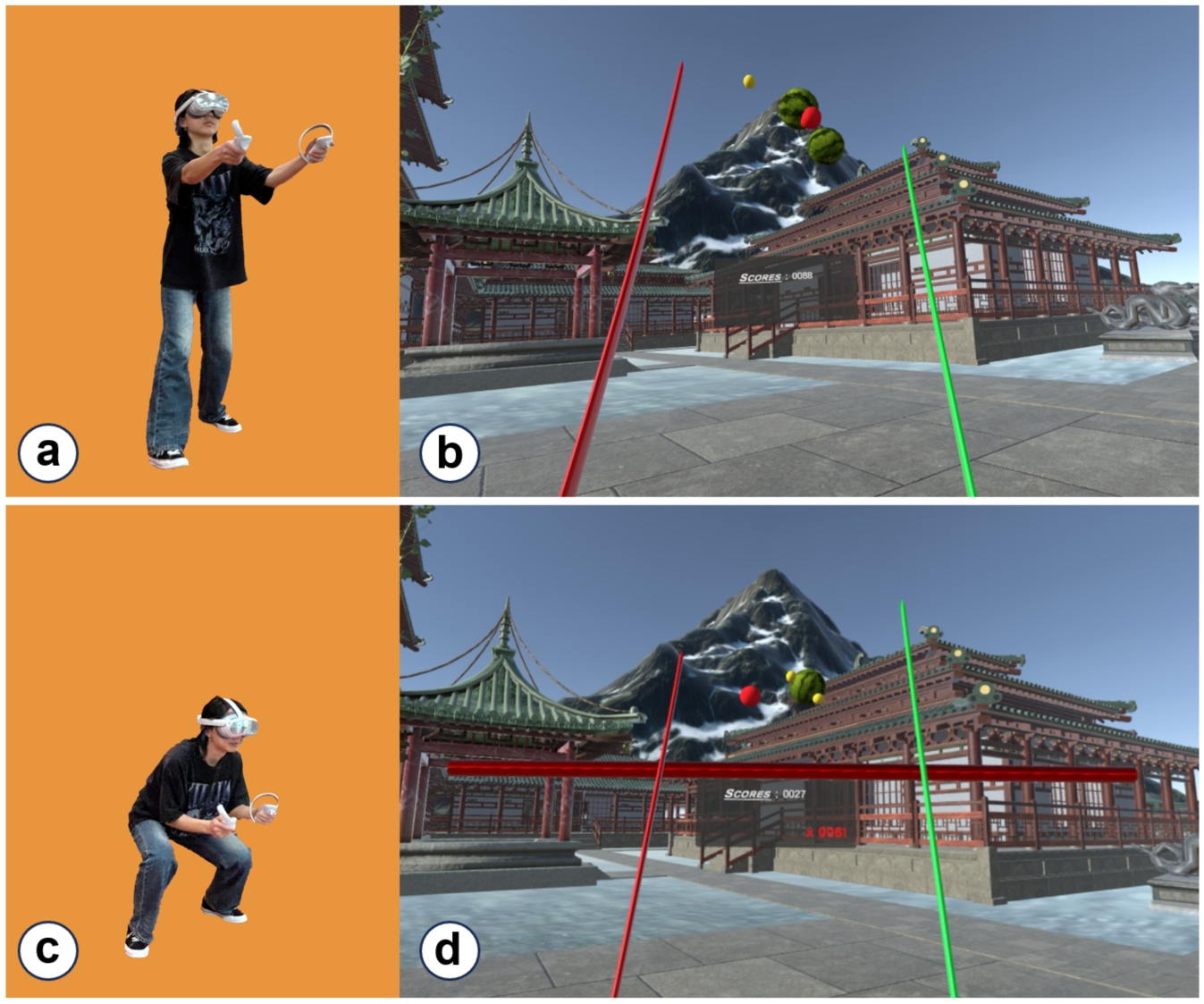}
    \caption{(a) The user swings the controllers to (b) slice fruits in the game. (c) The user squats to (d) dodge the horizontal bar obstacle in the game.}
    \label{gameplay}
\end{figure}

\subsubsection{Gameplay Mechanics}
Players are required to utilize two handheld controllers to wield virtual swords within the game (Figure \ref{gameplay}). Their main goal is to slice through as many fruits, including watermelons, apples, and lemons, as possible while dodging incoming bombs and obstacles. Fruits and bombs are launched from both the left and right sides of the player, following a parabolic trajectory that ensures they land within the player's controllable range, similar to the mechanism seen in VR Fruit Ninja.

In the game, players encounter a wooden horizontal bar obstacle (Figure \ref{gameplay}), which adapts dynamically to the height level of the player's eyes, ensuring accessibility for players of different heights. Moving at a consistent speed of 2-3 meters per second, the bar originates from the same location as the fruit launch point, progressively advancing toward the player. To evade collision with the obstacle, players must swiftly crouch down, followed by promptly resuming their upright position to continue gameplay.

\subsubsection{Game Structure and Scoring}
The game lasts 5 minutes and 12 seconds, divided into 5 one-minute sequences with 3-second rest intervals between two sequences. Within each one-minute sequence, players encounter 30 rapid 2-second rounds, featuring a mix of fruits and bombs. Rounds typically offer 2-4 fruits, with bombs appearing every 4-5 rounds. Furthermore, obstacles appear every 6 seconds in each one-minute sequence. 

Players earn scores in two categories: total score and combo count. Slicing fruits and successfully dodging obstacles contribute corresponding points to the player's total score, while missing fruits, slicing bombs, and colliding with obstacles result in deductions from the score. As for the combo count, each sliced fruit increments the count by 1. However, missing fruits or hitting bombs reset the combo count to 0. To keep track of their scores, players can view the scoring panel positioned directly in front of them. Additionally, in case of a collision with an obstacle, the panel briefly displays the word ``HIT" to alert the player visually.


\subsection{Outcome Measures}
\begin{itemize}

\item {\textbf{Performance.}} We gathered the following performance metrics: (1) game score; (2) success rate of slicing fruits, avoiding bombs and obstacles; and (3) maximum combo count. 

\item {\textbf{Exertion.}} Exertion levels were evaluated using three measures: (1) average heart rate (AvgHR\%), we assessed the intensity of physical activity using the percentage of the participant's age-predicted maximum heart rate (calculated by 211-0.64×age \cite{nes2013age}); (2) calories burned; and (3) the Borg Rating of Perceived Exertion (RPE) scale, which ranges from 6 to 20 \cite{borg1982psychophysical}.

\item {\textbf{Experience.}} We assessed participants' experiences during the experiment through subjective questionnaires, focusing on three aspects:
\begin{itemize}
    \item \emph{\textbf{Game Experience.}} We utilized four subscales, encompassing a total of 18 items sourced from the Player Experience of Need Satisfaction (PENS) scale \cite{ryan2006motivational} to evaluate participants' game experience. Using a 7-point Likert scale, participants rated their agreement with each item. The four subscales were: (1) Competence: participants' perception of their skills and abilities in the game. (2) Autonomy: participants' perception of experienced freedom and choice in the game. (3) Presence: the depth of engagement experienced by participants while playing. (4) Intuitive Controls: participants' perception of their actions translated into in-game actions.
    \item \emph{\textbf{Social Experience.}} We evaluated two aspects of participants' social experience: (1) Co-presence, assessed using 3 items derived from the ``Co-presence" subscale of the Networked Minds Social Presence Measure \cite{harms2004internal}, rated on a 7-point Likert scale. (2) Social acceptability, measured using two items adapted from \cite{koelle2020social}. Participants were asked to rate their feelings regarding playing the VR exergame in the current environment. Responses were provided on two scales, one ranging from 1 (embarrassed) to 6 (comfortable), and the other from 1 (annoyed) to 6 (enjoyable).
    \item \emph{\textbf{Cybersickness.}} We used the Simulator Sickness Questionnaire (SSQ) \cite{kennedy1993simulator} to assess cybersickness. This questionnaire consists of 16 items rated on a scale from 0 (none) to 3 (severe), evaluating nausea, oculomotor discomfort, and disorientation. A total SSQ score exceeding 40 indicates an unsatisfactory simulator experience \cite{caserman2021cybersickness}.
\end{itemize}

\item {\textbf{Self-consciousness.}} Before commencing the experiment, we utilized the Self-consciousness Scale (SCS) \cite{fenigstein1975public} to assess participants' self-consciousness, consisting of a total of 23 items. Each item was rated on a scale ranging from 0 (extremely uncharacteristic) to 4 (extremely characteristic).

\item {\textbf{Ranking and Interview.}} Following the gaming sessions, participants were asked to rank the three conditions based on their experiences and provide detailed reasons for their rankings. Additionally, they were asked about their willingness to play VR exergames in public environments in the future. All interviews were audio-recorded and transcribed for subsequent analysis.


\end{itemize}
\subsection{Participants}
We enrolled a total of 18 participants (10 females; 8 males) with an average age of 23.8 years (SD = 2.27, range = 19 to 28) using a university social media platform. Among them, 8 were university staff or researchers, and 10 were university students. For sedentary behavior, 12 participants reported sitting for work or studying for more than 6 hours per day on workdays. Regarding their regular physical activity habits, 4 participants engaged in physical activity for more than 3 hours per week, 6 for 1-3 hours per week, and 8 engaged in less than 1 hour of physical activity per week. 13 of these participants reported prior experience with VR, with 4 using them on a weekly basis. 15 participants had previous exposure to exergames, with only 1 reporting regular weekly play. All participants volunteered for the study without receiving compensation.

\subsection{Procedure}
Participants were first introduced to the experiment's objectives and procedure in a small closed room. They were informed that the study would take place in a real-world public setting, specifically, the underground passageway on the university campus, with an experimenter present to ensure their safety. Each participant was then given a consent form to review and sign.

Before starting the experiment, participants completed a pre-experiment questionnaire, which included demographic information, SSQ \cite{kennedy1993simulator}, and SCS \cite{fenigstein1975public} scales. Participants then entered personal details (age, gender, height, and weight) into the Polar Beat mobile application. Resting heart rate measurements were captured using the Polar OH1 HR monitor, with participants instructed to relax and remain motionless for over one minute.

To familiarize participants with the game and equipment, a 3-minute training phase without Passthrough functionality was conducted. Upon confirming participants' proficiency with the game mechanics and equipment, the experimenter guided them to begin the experiment in either the small closed room or the underground passageway. In each condition, participants were assisted with wearing the VR headset and Polar OH1 by the experimenter.

After each condition, participants filled out questionnaires to evaluate their exertion level \cite{borg1982psychophysical}, game experience \cite{ryan2006motivational}, social experience \cite{harms2004internal, koelle2020social}, and cybersickness \cite{kennedy1993simulator}. Participants were given a rest period until they felt ready to proceed to the next condition, allowing their heart rates to return to resting levels. At the end of the experiment, participants participated in a semi-structured interview where they ranked the experimental conditions and provided qualitative feedback on their experiences. Each experimental session lasted approximately 40 minutes per participant.

\section{Results}
We first assessed the normality of the data using Shapiro-Wilk tests and Q-Q plots. For data that did not follow a normal distribution, we applied transformations using the Aligned Rank Transform (ART) \cite{wobbrock2011aligned}. We then conducted one-way repeated-measures ANOVAs (RM-ANOVA) and adjusted for multiple comparisons using Bonferroni corrections. In cases where Mauchly’s test indicated violation of the assumption of sphericity, Greenhouse-Geisser estimates were used to adjust degrees of freedom. Additionally, we conducted Pearson's bivariate correlation analyses to explore the correlations between participants' self-consciousness and their performance, exertion, and experience.

\begin{figure*}[htbp]
    \centering
    \includegraphics[width=\textwidth]{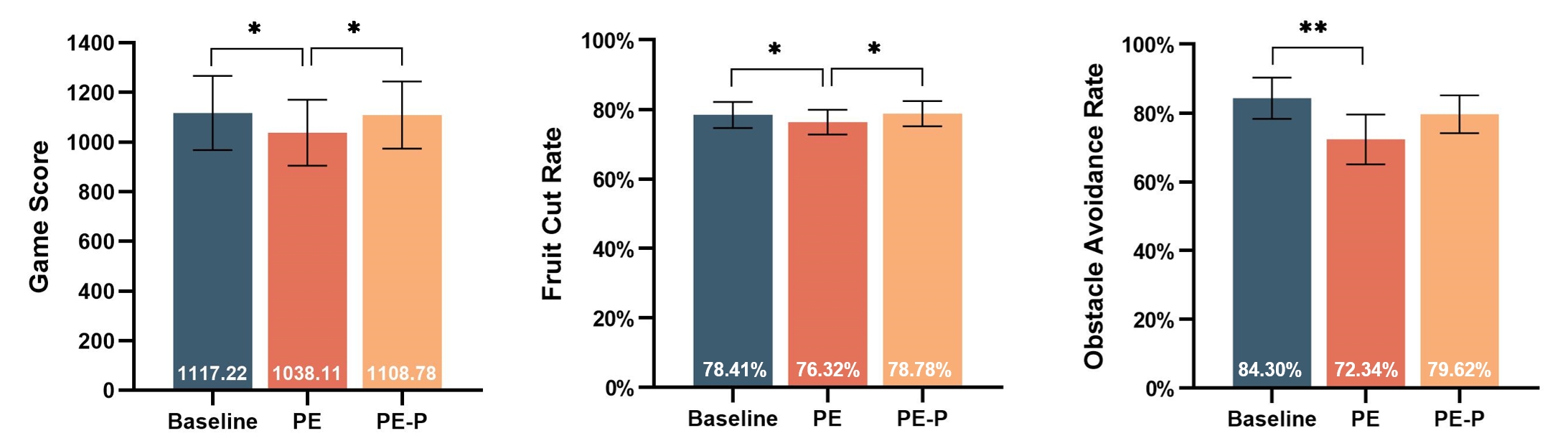}
    \caption{Mean game score, fruit cut rate, and obstacle avoidance rate for Baseline, PE, and PE-P. Error bars indicate 95\% confidence intervals. * and ** indicate statistical significance at the $p < 0.05$ and $p < 0.01$ levels, respectively.}
    \label{Performance}
\end{figure*}

\subsection{Performance}
As shown in Figure \ref{Performance}, significant differences in game scores between conditions were found ($F_{2,34}=6.735, p=.003, \eta_p^2=0.284$). Post-hoc tests revealed that participants achieved higher game scores in both the Baseline ($M=1117.22, SD=70.95$) ($p=.020$) and PE-P ($M=1108.78, SD=63.88$) ($p=.027$) conditions compared to the PE condition ($M=1038.11, SD=62.92$).

Furthermore, we found significant differences between conditions in the fruit cut rate ($F_{2,34}=8.443, p=.001, \eta_p^2=0.332$). The post-hoc tests indicated that participants achieved higher fruit cut rates in both the Baseline ($M=78.41\%, SD=0.02$) ($p=.015$) and PE-P ($M=78.78\%, SD=0.02$) ($p=.013$) conditions compared to the PE condition ($M=76.32\%, SD=0.02$).

The data for the obstacle avoidance rate underwent an ART due to non-normal distribution prior to conducting the RM-ANOVA. Significant differences between conditions were found ($F_{2,34}=5.642, p=.008, \eta_p^2=0.249$), with participants achieving a higher obstacle avoidance rate in the Baseline condition ($M=84.30\%, SD=0.12$) compared to the PE condition ($M=72.34\%, SD=0.15$) ($p=.007$). We did not find any significant effects for bomb avoidance rate, maximum combo count, or any significant correlation between participants' self-consciousness and their game performance.

\subsection{Exertion}
There were no statistically significant effects found regarding AvgHR\%, calories burned, and perceived exertion via Borg RPE 6-20. For Baseline, the mean AvgHR\% was 55.52\% ($SD=0.06$), mean calories burned were 30.94 ($SD=12.98$), and mean perceived exertion was 4.94 ($SD=1.55$). For PE, the mean AvgHR\% was 54.95\% ($SD=0.05$), with mean calories burned at 30.94 ($SD=12.98$), and mean perceived exertion recorded as 4.94 ($SD=1.55$). For PE-P, the mean AvgHR\% was 55.62\% ($SD=0.05$), mean calories burned were 30.44 ($SD=11.87$), and mean perceived exertion was 5.22 ($SD=1.90$).

Regarding correlation, as participants' self-consciousness increased, there was a significant increase in their avgHR\% in the PE condition ($r=.478, p=.045$). Apart from heart rate, no significant correlations were found in other measures.

\subsection{Experience}
\subsubsection{Game Experience}
\begin{figure*}[htbp]
    \centering
    \includegraphics[width=\textwidth]{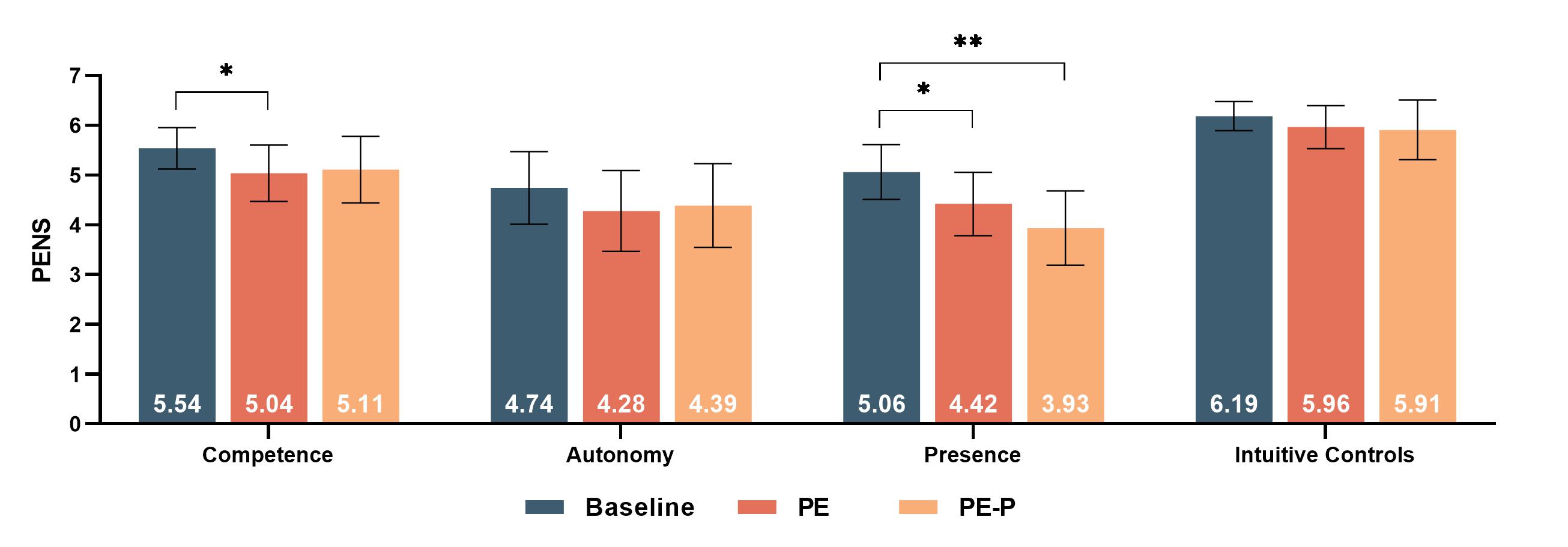}
    \caption{PENS ratings for Baseline, PE, and PE-P. Error bars indicate 95\% confidence intervals. * and ** indicate statistical significance at the $p < 0.05$ and $p < 0.01$ levels, respectively.}
    \label{pens}
\end{figure*}

The ratings for PENS in each condition are shown in Figure \ref{pens}. We noted significant differences between conditions on participants' perceived Competence ($F_{1.450, 24.656}=3.891, p=.046, \eta_p^2=0.186$). Further analysis using post-hoc tests indicated that participants reported a heightened sense of Competence in the Baseline condition ($M=5.54, SD=0.83$) compared to the PE condition ($M=5.04, SD=1.14$) ($p=.012$).

Concerning Presence, significant differences between conditions were found ($F_{1.374,23.356}=11.254, p=.001, \eta_p^2=0.398$). Post-hoc comparisons revealed that participants reported a higher level of Presence in the Baseline condition ($M=5.06, SD=0.26$) compared to the PE ($M=4.42, SD=0.30$) ($p=.012$) and PE-P ($M=3.93, SD=0.35$) ($p=.006$). No significant effects were found for Autonomy and Intuitive Controls.

Pearson's correlation analysis unveiled significant negative correlations between participants' self-consciousness and their perceived Competence in the PE condition ($r=-.517, p=.028$). Similarly, as self-consciousness levels rose, there was a significant decline in participants' perceived Presence in the public environments: PE ($r=-.540, p=.021$) and PE-P ($r=-.646, p=.004$). Additionally, with an increase in self-consciousness, participants reported a significant decrease in their perceived Intuitive Controls in the PE condition ($r=-.514, p=.029$). No significant correlations were found in Autonomy.

\subsubsection{Social Experience}
\begin{figure}[htbp]
    \centering
    \includegraphics[width=\columnwidth]{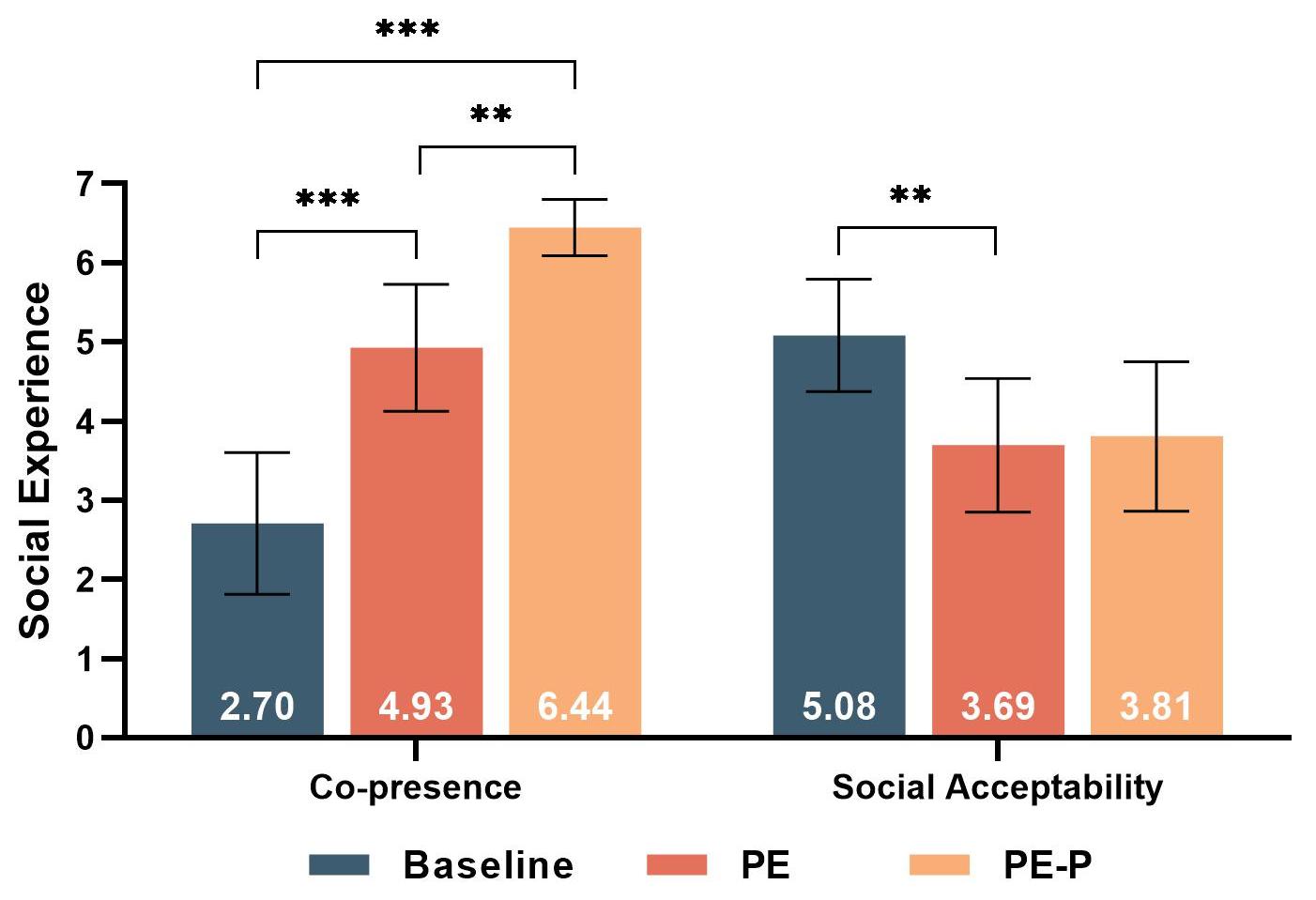}
    \caption{Co-presence and Social Acceptability ratings for Baseline, PE, and PE-P. Error bars indicate 95\% confidence intervals. *, **, and *** indicate statistical significance at the $p < 0.05, p < 0.01,$ and $p < 0.001$ levels, respectively.}
    \label{ps}
\end{figure}

Figure \ref{ps} displays the data of co-presence and social acceptability across all conditions. Significant differences were found concerning participants' perceived co-presence with others ($F_{2,34}=54.177, p=.000, \eta_p^2=0.761$). Participants reported a higher level of co-presence when in the PE-P condition ($M=6.44, SD=0.17$) compared to both the Baseline ($M=2.70, SD=0.42$) ($p=.000$) and PE ($M=4.93, SD=0.38$) ($p=.001$) conditions. Furthermore, participants reported a higher level of co-presence when in the PE condition compared to the Baseline condition ($p=.000$).

Concerning participants' social acceptability, significant effects were found ($F_{2,34}=7.048, p=.003, \eta_p^2=0.293$). Post-hoc comparisons revealed that participants reported a higher level of social acceptability in the Baseline condition ($M=5.08, SD=1.43$) compared to the PE condition ($M=3.69, SD=1.70$) ($p=.002$).

As participants' self-consciousness increased, we found a significant decrease in their perceived co-presence in the Baseline condition ($r=.522, p=.026$). Additionally, as self-consciousness levels rose, there was a significant decline in participants' perceived social acceptability in the PE condition ($r=-.591, p=.001$). 

\subsubsection{Cybersickness}
 Statistical analysis revealed no significant findings for total SSQ scores, Nausea, Oculomotor, and Disorientation. Across all three conditions—Baseline ($M=2.91, SD=3.53$), PE ($M=3.12, SD=3.90$), and PE-P ($M=3.12, SD=3.68$)—none of the participants exhibited total SSQ scores surpassing 20. For Baseline, the mean Nausea score was 4.77 ($SD=5.90$), mean Oculomotor score was 1.26 ($SD=2.91$), and mean Disorientation score was 1.55 ($SD=6.56$). For PE, the mean Nausea score was 4.24 ($SD=5.87$), mean Oculomotor score was 1.26 ($SD=2.91$), and mean Disorientation score was 3.09 ($SD=6.56$). For PE-P, the mean Nausea score was 3.71 ($SD=5.80$), mean Oculomotor score was 1.68 ($SD=3.24$), and mean Disorientation score was 3.09 ($SD=5.95$). 

\subsection{Ranking and Interview Results}
In terms of the ranking results, playing in a closed room was generally preferred by participants. 15 participants ranked the Baseline condition as their top choice, with 2 participants selecting it as their second choice. PE-P emerged as the second most preferred condition, garnering 2 first-choice and 10 second-choice rankings. On the other hand, PE exhibited the least favorable performance, receiving 1 first-choice ranking and 6 second-choice rankings.

When considering their choices, 9 participants highlighted that playing in the closed rooms was \emph{``quieter,"} \emph{``more immersive,"} and \emph{``more comfortable."} As for playing in public environments, although 3 participants indicated that \emph{``being observed by others could enhance motivation and performance,"} 8 participants perceived the environment as \emph{``chaotic and disruptive to gaming,"} 6 participants expressed a sense of \emph{``lack of security,"} and 5 participants mentioned feeling \emph{``awkward being watched by others."}

Nevertheless, 10 participants believed that Passthrough played a positive role in public environments, primarily because \emph{``it allows seeing the surroundings, increasing the sense of security."} Specifically, 6 participants noted the ability to \emph{``observe others' reactions,"} 4 participants appreciated \emph{``not having to worry about bumping into others,"} and 3 participants emphasized feeling \emph{``more comfortable when realizing that others are not very concerned about what I am doing."} Although most passersby simply passed by without much interaction, occasional individuals attempted to engage with the participants, such as by greeting them or inquiring about their activities. P7, P8, and P12 encountered such scenarios and emphasized the significance of being able to see others in such moments, as it \emph{``facilitates better communication"} and \emph{``reduces the likelihood of sudden surprises."} In contrast, 3 participants felt that using Passthrough in public environments \emph{``overemphasized the real world"}, while 2 participants expressed concerns that \emph{``being overly focused on others might lead to distractions."}

Regarding the willingness to play VR exergames in public environments, 9 participants expressed acceptance, 4 remained neutral, and 5 expressed refusal. 10 participants stated that playing VR exergames is \emph{``fun"} and \emph{``beneficial for physical health."} 8 participants believed that \emph{``the ability to use Passthrough is necessary in public environments."} Among the accepting participants, 4 stated that \emph{``I don't need to consider disturbing others because it's a public environment,"} while 3 stated that \emph{``there isn't much difference between playing in public environments and closed rooms."} The 2 neutral participants indicated that \emph{``I might accept public environments with fewer people."} As for the participants who were reluctant, their main concerns were \emph{``feeling awkward playing in public environments"} and \emph{``disliking exercising in public."}

\section{Discussion}
\subsection{Impact of Public Environments and Passthrough}
Our results demonstrate the impact of Passthrough on users' performance and experience in VR exergames in public environments. Users performed better in both the Baseline and PE-P conditions compared to the PE condition. Public environments often include the presence of other people, which can affect an individual's task performance. According to Social Facilitation Theory \cite{zajonc1965social, bond1983social}, an individual's performance can be enhanced or inhibited by the presence of others. This has been supported in VR research \cite{rettinger2022you}, where a co-located bystander led to social inhibition compared to a no-bystander condition. Our findings indicate social inhibition effects in the PE condition but no social facilitation effect in the PE-P condition.

One plausible explanation for the observed effects in the PE condition is rooted in the distraction-conflict theory \cite{BARON19861, SANDERS1981227}. This theory suggests that balancing task concentration with concerns about others' reactions or potential threats creates attentional conflict, leading to diminished performance. Participants in the PE condition were likely distracted by the fear of being judged by passersby or the risk of collisions. Insights from participant interviews support this notion, as they reported finding it challenging to concentrate due to uncertainty about whether passersby were observing them or if collisions were imminent, resulting in diminished task performance. In contrast, in the PE-P condition with Passthrough functionality, participants felt assured of their safety and understood that passersby were focused on navigating the passageway rather than observing them.


In terms of experience, we found a downgraded score in the presence rating of both public environment conditions, likely due to the noise of passersby \cite{eghbali2019social}. Despite prior research \cite{guo2024breaking, o2023re, mcgill2015dose} suggesting Passthrough might diminish gaming immersion, we did not confirm this, revealing distinctions between laboratory and public environments. Regarding social experience, participants had lower social acceptability when playing in the public environment without Passthrough. Interviews suggest this was due to security concerns, paralleling the ``sense of safety" highlighted by Eghbali et al. \cite{eghbali2019social}. Additionally, Passthrough notably bolstered users' sense of co-presence with passersby because the VR players can see the passersby.  


In summary, we found that when utilizing full virtual views, users' performance and experiences in public environments are indeed influenced, primarily reflected in poorer game performance, as well as lower perceived competence, presence, and social acceptance. Enabling the Passthrough feature resulted in better performance than without this feature. Furthermore, playing in public environments with Passthrough enhanced co-presence compared to without Passthrough and the Baseline condition.

\subsection{Individual Differences in User Performance and Experience Induced by Self-consciousness}
We found that playing VR exergames in public spaces without Passthrough poses significant challenges for individuals with high self-consciousness. This was reflected in their significantly increased heart rate, decreased perceived competence, presence, and sense of control, as well as reduced social acceptance when in the PE condition. Woods et al. \cite{woods2022impact} focused on the willingness and anxiety of using VR in public spaces, finding that extrovert individuals are more willing to try VR and experience lower anxiety when surrounded by a larger number of bystanders, while introvert individuals exhibit the opposite behavior. In contrast, our study focused on performance and experiences during the usage process, revealing the tension and vulnerability experienced by individuals with high self-consciousness in public environments, although this did not lead to a decrease in their performance.

Furthermore, while most of the correlations were found in the PE condition, we did not find any other significant correlations in the PE-P condition besides Presence. This indicated that using Passthrough in public environments might have a positive impact on improving the experience of users with high self-consciousness. These users are often hesitant to use VR in public environments due to concerns about disconnecting from reality, as well as worries about bystanders' opinions \cite{williamson2019planevr}. Therefore, being able to see the reactions of other passersby is important for them, which is consistent with the participants' interviews. Our results demonstrated that Passthrough can enhance the visibility of the surrounding environment, especially passersby, for users with high self-consciousness, thereby helping them overcome the challenges posed by public environments.

\subsection{Practical Implications}
Past research \cite{yoo2020embedding, touloudi2022applicability, stranick2021virtual, xu2021effects, sousa2022active, liu2021effects} has validated the physical and emotional benefits of VR exergames for office workers and students. Our study demonstrates the feasibility of integrating VR exergames into real-life work or study environments. Setting up a closed room in public spaces is optimal for VR exergames when feasible. However, when this is not possible, allowing sedentary users a convenient and safe way to play VR exergames in public environments is viable if the Passthrough feature is provided in VR headsets. Most participants believe that environmental awareness is crucial in public environments, helping them observe their surroundings and others' reactions.

Moreover, due to the continuous flow of people in public environments, maintaining constant environmental awareness is necessary. Previous research \cite{o2023re, von2019you} in small shared spaces, like living rooms, suggests providing users with brief visual cues when bystanders interact with them. However, our study found that in public environments, constant environmental awareness helps alleviate tension and anxiety, makes users feel safer, avoids surprises, and supports better game enjoyment.

Furthermore, many users appreciate the benefits of VR exergames and hold positive attitudes towards their future use in public environments, primarily due to the open and free nature of such environments. However, researchers and designers should still acknowledge individual traits in deploying VR exergames publicly, as some participants do not accept playing VR in public environments.


\subsection{Limitations and Future Work}
While our study sheds light on Passthrough's impact on VR exergame participation in public settings, several limitations should be acknowledged. First, the participants in our study were predominantly young adults. To generalize our findings across different age groups, future research could include participants from diverse demographic backgrounds, allowing for a more comprehensive understanding of how individuals of varying ages perceive and interact with VR technology in public settings.

Furthermore, our study involved a short-term experiment to explore the impact of the Passthrough feature. Although it demonstrated the potential to help users adapt better to public environments, its effects on promoting long-term usage and enhancing users' motivation for physical activities in these work and campus settings remain unknown. Future research could conduct long-term experiments to further investigate this aspect. 

Additionally, our study was conducted in an underground passageway, providing valuable insights into VR usage in such environments. To broaden the scope of our findings, future research could expand its investigations to include various other public environments and use other applications. Exploring diverse environments and application domains would deepen our understanding of how contextual factors influence user experiences and perceptions of VR technology in public settings.

\section{Conclusion}
This work represents a first-of-its-kind attempt to explore the impact of Passthrough on user performance and experience in real-world public environments, demonstrating its potential to help users incorporate VR exergames into their daily activities, thereby promoting healthy behaviors. We found that, compared to playing VR exergame in a closed space, users exhibit decreased performance in public environments if Passthrough functionality is not provided. Passthrough can enhance users' social acceptance to some extent in public environments without significantly compromising presence. Moreover, individual differences, especially self-consciousness, influenced user experiences, with Passthrough positively impacting users with higher self-consciousness by enhancing environmental awareness and social acceptance. Furthermore, our results contribute to understanding VR in public environments and underscore the potential of Passthrough technology to facilitate user adoption of VR in real-world environments.

\section*{Acknowledgment}
We thank the participants who joined our study. We also thank the reviewers for their insightful comments and helpful suggestions that helped improve our paper.

\bibliographystyle{abbrv-doi}

\bibliography{template}
\end{document}